\documentclass[12pt,preprint]{aastex}

\begin{document}


\title{Drift effects and the cosmic ray density gradient in a solar rotation period: First observation with
the Global Muon Detector Network (GMDN) \\}


\author{Y. Okazaki\altaffilmark{1}, A. Fushishita\altaffilmark{2}, T. Narumi\altaffilmark{2}, C. Kato\altaffilmark{2}, S. Yasue\altaffilmark{2}, T. Kuwabara\altaffilmark{3}, J. W. Bieber\altaffilmark{3}, P. Evenson\altaffilmark{3}, M. R. Da Silva\altaffilmark{4}, A. Dal Lago\altaffilmark{4}, N. J. Schuch\altaffilmark{5}, Z. Fujii\altaffilmark{6}, M. L. Duldig\altaffilmark{7}, J. E. Humble\altaffilmark{8}, I. Sabbah\altaffilmark{9}, J. K\'ota \altaffilmark{10}, and K. Munakata\altaffilmark{2}}


\altaffiltext{1}{Department of Geophysics, Tohoku University, Sendai, Miyagi 980-0861, Japan; okazaki@pat.geophys.tohoku.ac.jp}

\altaffiltext{2}{Physics Department, Shinshu University, Matsumoto, Nagano 390-8621, Japan; kmuna00@gipac.shinshu-u.ac.jp}

\altaffiltext{3}{Bartol Research Institute and Department of Physics and Astronomy, University of Delaware, Newark, DE 19716, USA; jwbieber@bartol.udel.edu}
\altaffiltext{4}{National Institute for Space Research (INPE), 12227-010 S\~{a}o Jose dos Campos, SP, Brazil; dallago@dge.inpe.br}

\altaffiltext{5}{Southern Regional Space Research Center (CRS/INPE), P.O. Box 5021, 97110-970, Santa Maria, RS, Brazil; njschuch@lacesm.ufsm.br}

\altaffiltext{6}{Solar Terrestrial Environment Laboratory, Nagoya University, Nagoya, Aichi 464-8601, Japan; fujii@stelab.nagoya-u.ac.jp}

\altaffiltext{7}{Australian Antarctic Division, Kingston, Tasmania 7050, Australia; marc.duldig@aad.gov.au}

\altaffiltext{8}{School of Mathematics and Physics, University of Tasmania, Hobart, Tasmania 7001, Australia; john.humble@utas.edu.au}

\altaffiltext{9}{Department of Physics, Faculty of Science, Kuwait University, Kuwait, on leave from Physics Department, Faculty of Science, University of Alexandria, Egypt; sabbahsom@yahoo.com}

\altaffiltext{10}{Lunar and Planetary Laboratory, University of Arizona, Tucson, AZ 87721, USA; kota@LPL.Arizona.EDU}


\begin{abstract}
We present for the first time hourly variations of the spatial density gradient of 50 GeV cosmic rays within a sample solar rotation period in 2006. By inversely solving the
transport equation, including diffusion, we deduce the gradient from the anisotropy that is derived from the observation made by the Global Muon Detector Network (GMDN). The anisotropy
obtained by applying a new analysis method to the GMDN data is precise and free from atmospheric temperature effects on the muon count rate recorded by ground based detectors. We find
the derived north-south gradient perpendicular to the ecliptic plane is oriented toward the Helioshperic Current Sheet (HCS) (i.e. southward in the toward sector of the
Interplanetary Magnetic Field (IMF) and northward in the away sector). The orientation of the gradient component parallel to the ecliptic plane remains similar in both sectors
with an enhancement of its magnitude seen after the Earth crosses the HCS. These temporal features are interpreted in terms of a local maximum of the cosmic ray density at the
HCS. This is consistent with the prediction of the drift model for the $A<0$ epoch. By comparing the observed gradient with the numerical prediction of a simple drift model, we
conclude that particle drifts in the large-scale magnetic field play an important role in organizing the density gradient, at least in the present $A<0$ epoch. We also found that
corotating interaction regions did not have such a notable effect. Observations with the GMDN provide us with a new tool for investigating cosmic ray transport in the IMF.
\end{abstract}



\keywords{drift effects on cosmic ray transport --- cosmic ray density gradient --- heliospheric current sheet --- corotating interaction region --- global muon detector network}


\section{INTRODUCTION}

The drift model of cosmic ray transport in the heliosphere predicts a bidirectional latitude gradient of the galactic cosmic ray (GCR) density, pointing in opposite directions on
opposite sides of the heliospheric current sheet (HCS) \citep{Jok82}. The predicted spatial distribution of the GCR density has a minimum on the HCS in the ``positive" polarity
period of the solar polar magnetic field (also referred as $A>0$ epoch), when the interplanetary magnetic field (IMF) directs away from (toward) the Sun in the northern
(southern) hemisphere, while the distribution has the maximum on the HCS in the ``negative" period ($A<0$ epoch) with the opposite field orientation in each hemisphere. The field
orientation reverses every 11 years around the maximum period of solar activity. By analyzing the GCR density measured with the anti-coincidence guard on board {\it{ACE}} and
{\it{Helios}} satellites and by ground based neutron monitors (NMs), \cite{Ric96} and \cite{Ric04} investigated GCR modulation by Corotating Interaction Regions (CIRs),
which are formed at the leading edges of corotating high-speed solar wind streams originating in coronal holes on the solar surface. The anti-coincidence guards measure sub-GeV
GCRs, while the NMs have a median response to primary particles with energies of 10-20 GeV. The above analyses were based on a one dimensional spatial distribution of the GCR
density measured as the temporal variation of the GCR count rate recorded by each detector as it sweeps across the CIR. From statistical analyses of variations in many CIRs with
and without the HCS, the above authors concluded that the IMF sector boundary (i.e. HCS) does not organize the GCR density, contrary to expectations from the simple drift model.
On the other hand, they also found observational evidence that the amplitude of the CIR-related modulations of the GCR density around the solar minimum is $\sim$50 \% larger in
$A>0$ epochs than in $A<0$ epochs. The simple drift model also predicts different modulation amplitudes in $A>0$ and $A<0$ epochs. Despite opposite latitudinal gradients, the
longitudinal profile has the same sense; the intensity being higher at the sector crossing and lower away from the current sheet in both cases \citep{Jok82}. The magnitude of the
drift effect on the GCR modulation in CIRs is still an open question.

The spatial distribution of the GCR density can be also inferred from measuring the anisotropic streaming of GCRs at the Earth, as the diffusion streaming of GCRs reflects the
local spatial gradient of the GCR density. The anisotropic streaming of GCRs in three dimensions is manifested as the diurnal and ``north-south" (NS) anisotropies of the intensity
recorded by ground-based detectors. The diurnal anisotropy is the anisotropy component along the equatorial plane, while the NS anisotropy is the component along the Earth's spin
axis. By analyzing the yearly mean anisotropy observed with several NMs and muon detectors, a series of studies revealed long term variations of the radial and latitudinal
gradients of GCR density in the heliosphere as well as variations in mean-free-paths of GCRs scattered by magnetic irregularities \citep{Swi69, Bie91, Che93, Hal96, Hal97, Ahl97,
Sab99, Mun02}. These studies provide important information on the large-scale distribution of GCRs and enable us to make crucial tests of physical models for the modulation of
GCRs in the heliosphere. By analyzing the diurnal anisotropy observed with several NMs, \cite{Bie91} first confirmed the existence of the bidirectional latitude gradient predicted
by the drift model. The gradient reversed with the solar magnetic polarity reversal at approximately every solar sunspot maximum. Similar results were also reported from
observations with muon detectors monitoring the GCR intensity at higher energies \citep{Hal96, Hal97, Mun02}.

These studies, however, only reported the yearly mean diurnal anisotropy. \cite{Bie91} actually analyzed the average anisotropy over the 27-day solar rotation period but used 14
rotation averages in a year for evaluating the error of the yearly mean. No information is deduced about the temporal evolution of the density gradient over shorter periods within
each solar rotation, such as evolutions on day-to-day or hourly timescales. The diurnal variation of GCR intensity recorded in a single ground-based detector gives the correct
anisotropy only when the anisotropy is stationary at least for one day, during which the detector scans 360$\degr$ longitude in space. In reality, the anisotropic streaming in
three dimensions changes dynamically both in magnitude and orientation. Accurate analysis of such dynamic changes requires continuous monitoring of the GCR intensity over the
entire sky. This is important to differentiate the temporal variation of the anisotropy from the variation of the omni-directional intensity (i.e. the GCR density). Such
monitoring has been made possible by construction of the worldwide networks of NMs and muon detectors \citep{Bie98, Kuw04}. From a multi-station analysis of NM data, \cite{Bie98}
concluded that the ``${\bf B}\times {\bf G}$" drift, arising from the local gradient of GCR density ($ {\bf G} $) and the IMF ($ {\bf B} $), is a primary source of the anisotropy
related to the arrival of Coronal Mass Ejections (CMEs) at the Earth.

Following the conclusion by \cite{Bie98}, \cite{Kuw04} analyzed the anisotropy observed with a network of muon detectors and showed that the three dimensional GCR distribution in
a CME and its temporal evolution over 6 hours is reasonably consistent with a magnetic flux rope structure expanding and passing the Earth. Muon detectors measure high-energy
GCRs by detecting secondary muons produced from the hadronic interactions of primary GCRs (mostly protons) with atmospheric nuclei. Since relativistic muons have relatively long
lifetimes (the proper half-life is ~2.2 $\mu$s) and can reach the ground preserving the incident direction of the initiating primary particles, we can measure the GCR intensity in
various directions with a multi-directional detector at a single location. This is different from NM measurements where directional information is only determined by the
geomagnetic field at the time of the observation and the rigidity of the initiating primary.

The observed intensity of muons at ground level is subject to atmospheric temperature effects. Atmospheric expansion with increasing temperature causes an increase of the mean
altitude of muon production from the decay of parent pions.  The increased flight time results in the decay of more muons into electrons and neutrinos before they reach the
detector. This effect is known as the ``negative temperature effect" and causes a decrease (increase) of the muon count rate with increasing (decreasing) temperature. \cite{Kuw04}
eliminated the temperature effect by normalizing the 24-hour trailing moving average (TMA) of the muon intensity recorded in each directional channel to that in the vertical
channel of one of the muon detectors in the network. This method works well when we analyze the dynamically changing anisotropy within a day in association with the CME arrival at
Earth. However, as is shown later, it does not give correct results in the derivation of the north-south anisotropy and its relationship to the IMF sector structure.

\cite{Nag72} proposed an observational method to measure the NS anisotropy by taking the difference between muon intensities in the north- and south-viewing channels in a
multi-directional muon detector at Nagoya in Japan. By taking the difference the atmospheric temperature effects common to all directional channels are removed \citep{Sag86}. The
difference is called the ``GG component" (see Section 3.1, equation 10). The NS anisotropy represented by the average GG component has been regarded as arising from the drift
streaming expressed by $ {\bf B} \times {\bf{G}}_{r} $, where ($ {\bf{G}}_{r} $) is the radial density gradient. This streaming reverses in association with the reversal of $
{\bf{B}} $ when the Earth crosses the HCS, as $ {\bf{G}}_{r} $ remains pointing away from the sun on average \citep{Swi69}. The GG component is regarded as a good indicator of
the IMF sector polarity. It has been reported, from a comparison between {\it{in-situ}} measurements of the IMF polarity and the sign of the GG component, that the IMF polarity
can be inferred from the GG component with an overall success rate exceeding 70 \% on a daily basis \citep{Mor79, Lau03}. In many cases, however, the GG component can only be used
after averaging over a one day period. This is due to the diurnal anisotropy also contributing to the GG component. The diurnal anisotropy, when observed by a detector on the
Earth, produces a diurnal variation in which the muon rate in each directional channel varies as a sinusoidal function of local time. Since the amplitude and local time of maximum
of this diurnal variation is generally different in different directional channels, the variation remains present even in the GG component. Only after averaging over one day does
the GG component become free of the contribution from the stationary diurnal anisotropy. Therefore, the GG component from a single muon detector at Nagoya cannot be used for
monitoring the NS anisotropy for periods shorter than one day.

The GCR modulation in the CIR and/or the IMF sector structure has so far been studied by analyzing the temporal variation of the GCR density, as the temporal variation is expected
to reflect the spatial distribution of the density along the detector's viewing path in the solar wind. A single detector measurement of the density cannot, however, observe the
spatial distribution separately from the temporal variation of the density at a particular location in space. The spatial distribution sampled by a single detector also does not
reflect the global distribution surrounding the detector, above and below the detector's viewing path. It is also affected by local irregularities in the solar wind. This can be
serious, particularly when analyzing satellite measurements of low energy (sub-GeV) GCRs. In this paper, we study the GCR modulation in CIRs for the first time by analyzing the
spatial gradient of the high-energy GCR density in three dimensions. The gradient gives information on the remote distribution surrounding the detector view.

We derive the density gradient from the GCR anisotropy observed with the Global Muon Detector Network (GMDN), which began operation in March 2006. The GMDN has been designed and
constructed to precisely measure the anisotropy. The responses of different surface muon detectors to GCRs have median energy ($\sim$ 50-150 GeV). These high-energy particles have
large Larmor radii in the IMF ($\sim$ 0.2 AU at 5 nT) and are less sensitive to small-scale irregular structures in the solar wind. Observation using the GMDN are, therefore,
expected to provide us with a new tool for investigating the large-scale structure in CIRs. Specifically, we develop a new analysis method by which we can eliminate the
atmospheric temperature effect on the muon count rate and precisely examine the temporal evolution of the NS anisotropy, even over a period shorter than a day. By applying this
method to data observed during CIRs in 2006 and 2007, we demonstrate that both the GMDN and the new analysis method enable us to monitor dynamic variations of the anisotropy in
association with sector boundary crossings by the Earth. We then compare the derived density gradient with the prediction of the drift model.

The outline of this paper is as follows. In section 2 we briefly describe the GMDN. In section 3.1 we first show how the temperature effect appears in the anisotropy derived from
the GMDN observations. In particular we demonstrate that the conventional method employed for eliminating the temperature effect diminishes the north-south anisotropy as well. We
then develop a new analysis method and apply it to the GMDN data. A separate analysis of the temperature effect using the high-altitude atmospheric data is also given in Appendix
A. This proves that the temperature effect was successfully eliminated by the new analysis method developed in section 3.1. In section 3.2, we examine the correlation between the
derived north-south anisotropy and the IMF sector polarity and confirm the performance of the new method. In section 4, we derive the spatial gradient of the cosmic ray density
from the best-fit anisotropy in three dimensions and compare it with numerical results obtained from a simple drift model. Our conclusions and discussions are given in section 5.

\section{GLOBAL MUON DETECTOR NETWORK (GMDN)}


Data acquisition by the network began in December 1992, as two-hemisphere observations using a pair of muon detectors at Nagoya (Japan) and Hobart (Australia), which have detection areas
of 36 m$^2$ and 9 m$^2$, respectively. Each of these detectors is multidirectional, allowing us to simultaneously record the intensities in 30 directions of viewing. Another small
(4 m$^2$) detector, prototype S\~{a}o Martinho (Brazil), was added to the network in March 2001 to fill a gap in directional coverage of the network over the Atlantic and Europe.
These detectors have identical design, except for their detection area, consisting of two horizontal layers of plastic scintillators, vertically separated by 1.73 m, with an
intermediate 5 cm layer of lead to absorb the soft component. Each layer comprises an array of 1 m$^2$ unit detectors, each with a 1 m$\times$1 m plastic scintillator viewed by a
photomultiplier tube of 12.7 cm diameter. By counting pulses of the 2-fold coincidences between a pair of detectors on the upper and lower layers, we can record the rate of muons
from the corresponding incident direction. The multi-directional muon telescope comprises various combinations between the upper and lower detectors. The prototype S\~{a}o
Martinho was upgraded in December 2005 by expanding its detection area to 28 m$^2$.

In March 2006, the Global Muon Detector Network (GMDN) was finally completed by installing a new detector at Kuwait University (Kuwait) with a detection area of 9 m$^2$.  This
adds new directions of viewing over the African Continent and the western Indian Ocean. The Kuwait University muon detector is a hodoscope designed specifically for measuring the
``loss cone" anisotropy, which is observed as a precursor to the arrival of interplanetary shocks at Earth and is characterized by an intensity deficit confined to a narrow pitch
angle region around the sunward IMF direction \citep{Mun00, Lee03}. Unlike the other three detectors, the Kuwait University detector consists of four horizontal layers of 30
proportional counter tubes (PCTs). Each PCT is a 5 m long cylinder with a 10 cm diameter having a 50-micron thick tungsten anode along the cylinder axis. A 5 cm layer of lead is
installed above the detector to absorb the soft component. The PCT axes are aligned geographic east-west (X) in the top and third layers and north-south (Y) in the
second and bottom layers. The top and second layers form an upper pair, while the third and bottom layers form a lower pair. The two pairs are separated vertically by 80 cm. 
Muon recording is triggered by the fourfold coincidence of pulses from all layers and the incident direction is identified from X-Y locations of the upper and lower PCT pairs.
This is approximately equivalent to recording muons with two 30$\times$30 square arrays of 10 cm$\times$10 cm detectors separated vertically by 80 cm. The muon count is
recorded in each of 23$\times$23=529 directional channels which cover 360$\degr$ of azimuth angle and 0$\degr$ to 60$\degr$ zenith. For analyzing Kuwait University data
together with the data from the other three detectors of different geometry, we convert 529 directional channels into 13 channels, which are equivalent to those in Hobart having
the same detection area (9 m$^2$). For the performance of a muon hodoscope of similar design, readers can refer to \cite{Mun05}.

Table 1 summarizes the locations of detectors and characteristics of directional telescopes in the GMDN. The total number of directional channels available in the network was 60
at March 2006. The median rigidity ($ P_m $) of primary cosmic rays recorded, calculated by utilizing the response function of the atmospheric muons to the primary
particles \citep{Mur79}, ranges from 55 to 114 GV and the statistical error ($\sigma_c$) of hourly count rate ranges between 0.06 and 0.49 \%. Each symbol in Figure 1 shows the
asymptotic viewing direction, after correction for geomagnetic bending, at rigidity $ P_m $ of each directional channel, as determined using a particle trajectory code
\citep{Lin95}. The track through each symbol represents the spread of viewing directions for particle rigidities between $ P_{0.1} $ and $ P_{0.9} $ that confine the central 80 \%
of each directional channel's energy response. Figure 1 illustrates that the GMDN covers almost the entire sky, although it still has gaps remaining in its directional coverage
over North America and the southern Indian Ocean.




\section{DATA ANALYSIS METHOD AND IT'S PERFORMANCE}

\subsection{Derivation of the cosmic ray anisotropy in three dimensions} \label{bozomath}

In this section, we analyze pressure corrected hourly muon count rates recorded in the 60 directional channels in the GMDN between 2006 March 14 and 2007 March 4. This period
covers 13 solar rotations (Carrington Rotations CR2041-CR2053) during which clear recurrent signatures of CIRs were seen in solar wind parameters measured by the {\it{ACE}}
satellite. We chose this period because {\it{ACE-Level2}} data were available.\footnote{{\it{ACE-Level2}} data are available at http://www.srl.caltech.edu/ACE/ASC/} We derive the
anisotropy by fitting the function $ I^{fit}_{i,j}(t) $  given by
\begin{eqnarray}
I^{fit}_{i,j}(t) = I^{0}_{i,j}(t) + \xi^{GEO}_{x}(t) ( c^{1}_{1i,j} \cos \omega t_i - s^{1}_{1i,j} \sin \omega t_i )  \nonumber \\
 + \xi^{GEO}_{y}(t) ( s^{1}_{1i,j} \cos \omega t_i + c^{1}_{1i,j} \sin \omega t_i )  + \xi^{GEO}_{z}(t) c^{0}_{1i,j},
\end{eqnarray}
to the observed hourly count rate $I^{obs}_{i,j}(t)$ of muons at universal time $t$ in the $j$-th directional channel of the $i$-th muon detector in the GMDN \citep{Kuw04}. In
this equation, $\xi^{GEO}_{x}(t)$, $\xi^{GEO}_{y}(t)$, $\xi^{GEO}_{z}(t)$ are the best-fit parameters denoting three components of the anisotropy; $c^{1}_{1i,j}$, $s^{1}_{1i,j}$
and $c^{0}_{1i,j}$ are the coupling coefficients calculated by assuming a rigidity independent anisotropy; $t_i$ is the local time at the location of the $i$-th detector and
$\omega = \pi /12$. The anisotropy components in (1) are defined in a local geographical coordinate system (GEO), in which the $z$-axis is directed toward geographic north, the
$x$-axis is in the equatorial plane and directed to the zenith of a point on the Earth's equator at 00:00 local time and the $y$-axis completes the right-handed coordinate set. Note
that the anisotropy vector points toward a direction \it{from} \rm which the highest GCR flux is measured; i.e., the anisotropy vector is oppositely directed to the GCR streaming
vector. Among the three components of the anisotropy, $\xi^{GEO}_{z}(t)$ represents the north-south (NS) anisotropy. We assume $I^{0}_{i,j}(t)$ as
\begin{equation}
I^{0}_{i,j}(t) = I_{0}^{\prime}(t) + I_{0}^{\prime \prime}(t)P_{m i,j}/P_{m 1,1},
\end{equation}
where $I_{0}^{\prime}(t)$ and $I_{0}^{\prime \prime}(t)$ are best-fit parameters and $P_{m i,j}$ and $P_{m 1,1}$ are the median primary rigidities ($P_{m}$ in Table 1) of the
$j$-th directional channel of the $i$-th muon detector and Nagoya vertical channel ($i=1, j=1$), respectively. The first term on the right hand side represents the energy
independent portion of the density, while the second term represents the rigidity dependence. We define the best-fit density $I_{0}(t)$ as
\begin{equation}
I_{0}(t)=I_{0}^{\prime}(t)+I_{0}^{\prime \prime}(t),
\end{equation}
which is equal to $I^{0}_{1,1}(t)$ for the vertical channel of Nagoya with $P_{m}=59$ GV. We determine the best-fit set of parameters ($I_{0}(t)$, $\xi^{GEO}_{x}(t)$,
$\xi^{GEO}_{y}(t)$, $\xi^{GEO}_{z}(t)$) that minimize the residual $S$, defined as
\begin{equation}
S=\sum_{i,j}{(I^{obs}_{i,j}(t)-I^{fit}_{i,j}(t))^{2}/\sigma_{ci,j}^2},
\end{equation}
where $\sigma_{ci,j}$ is the count rate error in the $j$-th directional channel of the $i$-th muon detector.

We first fit (1) to the observed percentage deviation $r^{obs}_{i,j}(t)$ of the pressure-corrected hourly counting rate from the average over each solar rotation. Gray curves in Figures
2b-e display the best-fit density and three GEO components of the anisotropy, while Figure 2a shows the observed muon count rates in vertical channels at Nagoya (black curve) and
S\~{a}o Martinho (thin black curve), each as a function of the day of year (DOY) in a sample rotation period (CR2043) in 2006. We plot muon rates in two vertical channels to
demonstrate the atmospheric temperature effect in Figure 2a, but the parameters in Figures 2b-e are derived from best-fitting to data from all directional channels in the GMDN.

As illustrated by the gray curves in Figures 2b-e, best-fitting to the original percent deviation  $r^{obs}_{i,j}(t)$ results in large diurnal variations in the anisotropy,
particularly when the difference between the vertical muon rates at Nagoya and S\~{a}o Martinho becomes significant. This is due to the atmospheric temperature effect described in
Appendix A, simply because best-fitting the model intensity in (1) leads to an anisotropy from the direction with the higher count rate. For instance, as seen in Figure 2a in DOYs
140-144, the vertical muon rate is higher at S\~{a}o Martinho than at Nagoya, while the best-fit anisotropy component $\xi^{GEO}_{x}(t)$ in Figure 2c shows a maximum around 0300
UT when the viewing directions of the S\~{a}o Martinho muon detector point toward the GEO-$x$ direction. On the other hand, when the vertical muon rate is higher at Nagoya than at
S\~{a}o Martinho (e.g. DOY 148-151), the best-fit $\xi^{GEO}_{x}(t)$ shows a maximum around 1400 UT, corresponding to the time when the viewing directions at Nagoya point toward
GEO-$x$.

To avoid this spurious diurnal variation in the derived anisotropy, \cite{Kuw04} calculated $I^{obs}_{i,j}(t)$ for the best-fit by normalizing the 24-hour trailing moving average
(TMA) of $r^{obs}_{i,j}(t)$ to that of the vertical channel in the Nagoya muon detector as
\begin{equation}
I^{obs}_{i,j}(t) = r^{obs}_{i,j}(t) \bar{I}_{1,1}(t)/\bar{I}_{i,j}(t),
\end{equation}
where $ \bar{I}_{i,j}(t) $ is the 24-hour TMA of $ r_{i,j}(t) $ defined as
\begin{equation}
\bar{I}_{i,j}(t) = \sum_{t-23}^{t} r^{obs}_{i,j}(t)/24,
\end{equation}
and $ \bar{I}_{1,1}(t) $  is the TMA for the vertical channel at Nagoya. This normalization efficiently removes the spurious diurnal variation of the best-fit anisotropy and works
well for deriving the temporal variation over a period shorter than 24 hours. It has been used for analyzing the geometry of a Coronal Mass Ejection \citep{Kuw04}. The
normalization to the 24-hour TMA, however, does not work for the stationary NS anisotropy. It is clear from (1) that the stationary NS anisotropy $ \xi^{GEO}_{z} $ introduces no
temporal variation to $ I^{fit}_{i,j}(t) $, while $ \xi^{GEO}_{x} $ and $ \xi^{GEO}_{y} $ cause a diurnal variation through terms of $ \cos \omega t_i $ and $ \sin \omega t_i $.
If the NS anisotropy is stationary over a period longer than 24 hours it is therefore eliminated by the normalization of $ r^{obs}_{i,j}(t) $ in (5). We tested the normalization
to the 24-hour TMA which resulted in a poor correlation between $ \xi^{GEO}_{z} $ and IMF sector polarity, even during a period when the GG component shows a good correlation.

To overcome this difficulty, we develop a new analysis method which is free from both atmospheric temperature effects and the normalization to the TMA. In this new method we
obtain the anisotropy by best-fitting the model to the difference between the muon rate in each directional channel and the vertical channel for that detector. This subtraction
efficiently eliminates the temperature effects, which are common to all directional channels at least as a first order approximation (see Appendix A). We define the observed
difference $ \Delta I^{obs}_{i,j}(t) $ as
\begin{equation}
\Delta I^{obs}_{i,j}(t) = r^{obs}_{i,j}(t) - r^{obs}_{i,1}(t),
\end{equation}
where $ r^{obs}_{i,1}(t) $ is the observed percent deviation for the vertical channel ($j=1$) of the $i$-th detector. We fit to $ \Delta I^{obs}_{i,j}(t) $ the model function $
\Delta I^{fit}_{i,j}(t) $ given as
\begin{eqnarray}
\Delta I^{fit}_{i,j}(t) =
I_{0}^{\prime \prime}(t) \Delta P_{m i,j}
+ \xi^{GEO}_{x}(t) ( \Delta c^{1}_{1i,j} \cos \omega t_i - \Delta s^{1}_{1i,j} \sin \omega t_i )  \nonumber \\
+ \xi^{GEO}_{y}(t) ( \Delta s^{1}_{1i,j} \cos \omega t_i + \Delta c^{1}_{1i,j} \sin \omega t_i )  + \xi^{GEO}_{z}(t) \Delta c^{0}_{1i,j},
\end{eqnarray}
where $\Delta P_{m i,j}$, $\Delta c^{1}_{1i,j}$, $\Delta s^{1}_{1i,j}$ and $\Delta c^{0}_{1i,j}$ are calculated as
\begin{eqnarray}
\Delta P_{m i,j}    = (P_{m i,j} - P_{m i,1})/P_{m 1,1},    \nonumber \\
\Delta c^{1}_{1i,j} = c^{1}_{1i,j} - c^{1}_{1i,1},          \nonumber \\
\Delta s^{1}_{1i,j} = s^{1}_{1i,j} - s^{1}_{1i,1},          \nonumber \\
\Delta c^{0}_{1i,j} = c^{0}_{1i,j} - c^{0}_{1i,1},
\end{eqnarray}
with $P_{m i,1}$, $ c^{1}_{1i,1} $, $ s^{1}_{1i,1} $ and $ c^{0}_{1i,1} $ denoting the median primary rigidity and coupling coefficients for the vertical channel of the $i$-th
detector. This best-fit gives no direct information on the GCR density $I_{0}(t)$, as the rigidity independent density ($I_{0}^{\prime}(t)$ in (2)) does not appear in (8). To
obtain $I_{0}(t)$, therefore, we first correct the observed vertical muon rate at Nagoya for atmospheric temperature effects by using high-altitude atmospheric data (see
Appendix A). We then calculate $I_{0}(t)$ by subtracting best-fit {\boldmath{$\xi$}}$^{GEO}(t)$ terms in (1) from the temperature corrected vertical muon rate. Black curves in
Figures 2b-e show the best-fit density and anisotropy derived by this method. It is clear that the spurious diurnal variation of the anisotropy arising from temperature effects is
substantially removed. We also corrected the observed muon rate for temperature effects by using high-altitude atmospheric data available at the location closest to each muon
detector in the GMDN (see Appendix A) and confirmed that these results were the same as those obtained by our new method. This shows that the new method succeeds in eliminating
temperature effects. Note that we can only apply the correction in Appendix A for part of the entire period analyzed here because high-altitude atmospheric data were not available
simultaneously for all muon detectors at many times.

In order to represent the NS anisotropy intensity parallel to the Earth's rotation axis, a difference combination (GG) of directional count rates from the Nagoya telescope was
developed. It is defined as
\begin{equation}
r_{GG}(t) = ( r_{N2}(t) - r_{S2}(t) ) + ( r_{N2}(t) - r_{E2}(t) ).
\end{equation}
where $r_{XX}(t)$ is the percent deviation of the pressure-corrected muon rate in the directional channel $XX$ (=N2, S2, E2) of the Nagoya muon detector from the average over each
solar rotation period \citep{Nag72}. The physical characteristics of the directional channels N2, S2 and E2 are listed in Table 1. The GG component is free from atmospheric
temperature effects, which are common to all directional channels used in the right hand side of (10) and are removed by the subtractions \citep{Sag86}. The GG component in CR2043
is displayed in Figure 2f, where the fluctuation is filtered out by taking 5-hour central moving averages. In this panel one can see a significant diurnal variation in the GG
component superposed on the day-to-day variation (i.e. DOYs 150-155). This diurnal variation is due to the contribution from the diurnal anisotropy perpendicular to the Earth's
rotation axis; i.e., $\xi^{GEO}_{x}(t)$ and $\xi^{GEO}_{y}(t)$ in (1). The anisotropy component, when stationary and observed by a detector on the rotating Earth, produces a
diurnal variation in which the muon rate varies as a sinusoidal function of local time. Since the amplitude and local time of maximum of this diurnal variation are both
generally different in different directional channels, the diurnal variation remains in the GG component even after the subtractions in (10). Owing to this contribution from the
diurnal anisotropy, the GG component from a single muon detector at Nagoya cannot be used as an indicator of the NS anisotropy on hourly basis. In many cases, it has been used
only after averaging over a one day period \citep{Mor79,Lau03}. In contrast, by analyzing data from the muon detectors in the GMDN, we can derive the NS anisotropy separately from
the diurnal anisotropy. The day-to-day variation shown by the black curve in Figure 2e is in good agreement with that shown for the GG component in Figure 2f, implying that the new
method successfully retains the NS anisotropy recorded by the GG component. Since this method gives us the NS anisotropy free from the diurnal variation, it enables us to examine
the temporal variation within a day as well as the gradual day-to-day variation. In the next subsection, we examine the correlation of the derived NS anisotropy with the IMF
sector polarity, preparing for our detailed analysis of the density gradient.

\subsection{Correlation of the NS anisotropy with the IMF sector polarity} \label{bozomath}

We define the IMF sector polarity based on the hourly mean IMF components in the geocentric solar ecliptic coordinate system (GSE) in {\it{ACE-Level2}} data. In the following
analysis we lag the {\it{ACE}} IMF and solar wind data by one hour, as a rough correction for the solar wind transit time between {\it{ACE}} and the Earth. As the spiral angle of
the IMF measured near the Earth is, on average, about 45$\degr$, the {\it{toward}} (the Sun) and {\it{away}} (from the Sun) sectors are defined relative to a plane normal to the
45$\degr$ spiral in the ecliptic plane; i.e., in GSE coordinates the IMF is designated {\it{toward}} (T) if $B_{x}>B_{y}$ and {\it{away}} (A) if $B_{y}>B_{x}$. Figure 3 shows
histograms of the GG component and the best-fit $\xi^{GEO}_{z}$ in a period between 2006 March 14 and 2007 March 4. The upper two panels (a \& b) display hourly values, while the
lower two panels (c \& d) show daily mean values. Figure 3a shows that the hourly values of the GG component have extremely large dispersion, partly due to the contribution from
the diurnal anisotropy as mentioned in the preceding section.
Figures 3b and 3d display the histograms by the new analysis method. The dispersion in Figure 3b is much smaller than that of the GG component in Figure 3a. The T/A separation in Figure 3b is also significantly improved when compared with $\xi^{GEO}_{z}$ by the normalization using the 24-hour TMA.

\section{GCR MODULATION IN A CIR AND A COMPARISON WITH THE DRIFT MODEL}

\subsection{Derivation of the GCR density gradient from the anisotropy} \label{bozomath}


In this section, we derive the spatial gradient of the GCR density from the best-fit anisotropy using the new analysis method described in the preceding section. Aiming to examine
the global drift effect on the GCR modulation within a solar rotation period, we now concentrate on analyzing a typical sample rotation (CR2043). Statistical analyses of multiple
rotations will be presented in a separate paper.

We first transform the anisotropy components, $\xi^{GEO}_{x}(t)$, $\xi^{GEO}_{y}(t)$ and $\xi^{GEO}_{z}(t)$ in the GEO, to the geocentric solar ecliptic coordinate system (GSE).
The coordinate system used hereafter in this paper is GSE, unless otherwise noted. We then correct the transformed anisotropy {\boldmath{$\xi$}}$^{GSE}(t)$ for the solar wind
convection anisotropy and the Compton-Getting anisotropy arising from the Earth's 30 km/s orbital motion and obtain the diffusive anisotropy {\boldmath{$\xi$}}(t) as
\begin{equation}
{\mbox{\boldmath{$\xi$}}}(t)={\mbox{\boldmath{$\xi$}}}^{GSE}(t)+(2+\gamma)({\bf{V}}_{SW}(t)-{\bf{v}}_{E})/c,
\end{equation}
where $\gamma$ is the power law index of the GCR energy spectrum, which we set to 2.7, ${\bf{V}}_{SW}(t)$ is the solar wind velocity in the {\it{ACE-Level2}} data, $ {\bf{v}}_{E}
$ is the velocity of the Earth's orbital motion and $c$ is the speed of light.

Figures 4b-d show hourly values of the three GSE components of {\boldmath{$\xi$}}(t) in CR2043 while Figure 4a shows the best-fit density $I_{0}(t)$. The data in T (A) sectors
are displayed by small open (filled) circles and the identified times of the Earth's HCS crossings are indicated by vertical lines. The dispersion of each parameter due to the
counting error ($\sigma_c$ in Table 1) is evaluated on the basis of the numerical simulation of the best-fit to be 0.06 \% for $I_{0}(t)$, 0.05 \% for $\xi_{x}(t)$, 0.06 \% for
$\xi_{y}(t)$ and 0.07 \% for $\xi_{z}(t)$. Note that in Figure 4d $\xi_{z}(t)$ is positive in the T-sectors for the majority of hours, while it is negative in the A-sectors,
similar to Figure 3b in the preceding section. Such a feature is much harder to see in the hourly values of the GG component in Figure 4e due to statistical fluctuations (the
dispersion of the GG component due to the count rate error is estimated to be 0.27 \%). It is also noted that $\xi_{x}(t)$ and $\xi_{y}(t)$ in Figures 4b \& c remain respectively
negative and positive in most hours, regardless of the sector polarity. We derive the GCR density gradient in three dimensions from {\boldmath{$\xi$}}(t) in this figure.

We divide the observed anisotropy vector {\boldmath{$\xi$}}(t) into components parallel and perpendicular to the IMF as
\begin{equation}
{\mbox{\boldmath{$\xi$}}}(t)={\mbox{\boldmath{$\xi_\parallel$}}}(t)+{\mbox{\boldmath{$\xi_\perp$}}}(t).
\end{equation}
These components are given as
\begin{equation}
{\mbox{\boldmath{$\xi_\parallel$}}}(t)=R_{L}(t)\alpha_{\parallel}{\bf{G}}_{\parallel}(t), \\
\end{equation}
\begin{equation}
{\mbox{\boldmath{$\xi_\perp$}}}(t)=R_{L}(t)(\alpha_{\perp}{\bf{G}}_{\perp}(t)-{\bf{b}}(t)\times{\bf{G}}_{\perp}(t)), \\
\end{equation}
where ${\bf{G}}_{\parallel}(t)$ and ${\bf{G}}_{\perp}(t)$ are respectively the density gradient components parallel and perpendicular to the IMF, $R_{L}(t)$ is the particle's
effective Larmor radius, ${\bf{b}}(t)$ is the unit vector pointing in the direction of the IMF, and $\alpha_{\parallel}$ and $\alpha_{\perp}$ are the dimensionless mean free paths
($\lambda_{\parallel}(t)$ and $\lambda_{\perp}(t)$) of GCR scattering by magnetic irregularities, defined as
\begin{equation}
\alpha_{\parallel}=\lambda_{\parallel}(t)/R_{L}(t)=3\kappa_{\parallel}(t)/R_{L}(t)/c, \\
\end{equation}
\begin{equation}
\alpha_{\perp}=\lambda_{\perp}(t)/R_{L}(t)=3\kappa_{\perp}(t)/R_{L}(t)/c, \\
\end{equation}
where $\kappa_{\parallel}(t)$ and $\kappa_{\perp}(t)$ are respectively the parallel and perpendicular diffusion coeffcients. We use for $ R_{L}(t) $ the Larmor radius of 50 GeV
protons in the IMF magnitude $B(t)$ in the {\it{ACE-Level2}} data. The second term on the right-hand side of (14) expresses the drift anisotropy, while the first term represents
the perpendicular diffusion. The parallel gradient ${\bf{G}}_{\parallel}(t)$ in (13) can be directly derived from {\boldmath{$\xi_\parallel$}}(t) as
\begin{equation}
{\bf{G}}_{\parallel}(t)={\mbox{\boldmath{$\xi_\parallel$}}}(t)/R_{L}(t)/\alpha_{\parallel}.
\end{equation}
On the other hand, the equation (14) which applies only for the perpendicular component can be rewritten as
\begin{equation}
{\mbox{\boldmath{$\xi_\perp$}}}(t)=R_{L}(t) \left(
\begin{array}{ccc}
\alpha_{\perp} &  b_z(t)       & -b_y(t)       \\
-b_z(t)       & \alpha_{\perp} &  b_x(t)       \\
 b_y(t)       & -b_x(t)       & \alpha_{\perp} \\
\end{array}
\right){\bf{G}}_{\perp}(t).
\end{equation}
By solving (18) with respect to ${\bf{G}}_{\perp}(t)$, we get
\begin{equation}
{\bf{G}}_{\perp}(t)=
\left(
\begin{array}{ccc}
\alpha_{\perp} &  b_z(t)       & -b_y(t)       \\
-b_z(t)       & \alpha_{\perp} &  b_x(t)       \\
 b_y(t)       & -b_x(t)       & \alpha_{\perp} \\
\end{array}
\right)^{-1}{\mbox{\boldmath{$\xi_\perp$}}}(t)/R_{L}(t).
\end{equation}
This equation cannot be used for the case of $\alpha_{\perp}=0$, in which the inverse matrix in (19) becomes singular. 
In the case that cross-field diffusion is negligible relative to the drift streaming in (14), ${\bf G_{\perp}}(t)$ is given by
\begin{eqnarray}
{\bf G_{\perp}} (t) = {\bf b} (t) \times {\mbox{\boldmath{$\xi$}}} (t)/ R_{L}(t) =
\left(
\begin{array}{ccc}
0        &   -b_z(t)   &   b_y(t)   \\
b_z(t)  &   0        &   -b_x(t)    \\
-b_y(t)   &   b_x(t)  &   0         \\
\end{array}
\right)
{\mbox{\boldmath{$\xi$}}}(t)/R_{L}(t).
\end{eqnarray}
Ignoring the minor IMF component ($b_z$) perpendicular to the ecliptic plane, we find in (20) that the NS anisotropy ($\xi_{z}(t)$) in the GSE coordinate system reflects the
ecliptic component of the gradient ($G_{\perp x}(t)$ and $G_{\perp y}(t)$), while the anisotropy components in the ecliptic plane ($\xi_{x}(t)$ and $\xi_{y}(t)$) are indicators of
the NS gradient ($G_{\perp z}(t)$). In the next section, we adopt two combinations of $\alpha_\parallel$ and $\alpha_\perp$ to derive the gradient from the observed anisotropy. We
first adopt an ad-hoc choice of $\alpha_\parallel=7.2$ and $\alpha_\perp=0.36$, which will be used in the numerical calculation of a simple drift model. We then calculate
${\bf{G}}(t)$ with {\boldmath{$\xi_\parallel$}}=0 and $\alpha_\perp=0$ for the case of dominant drift streaming and compare the two.

\subsection{Density gradient in a sample rotation} \label{bozomath}

Figure 5 shows the GCR density gradient derived from the anisotropy in Figure 4 by (17) and (19) together with the solar wind velocity and the IMF data in a sample rotation,
CR2043. From top to bottom, each panel displays (a) the GSE longitude of the IMF, (b) the magnitude of the IMF, (c) the solar wind velocity and (d-f) the three GSE components of the
derived gradient vector. Open and filled circles in each panel display the hourly values in T and A sectors, respectively. The IMF sector boundaries are indicated by vertical
lines. A four-sector IMF structure is evident in this rotation. Enhancements of the IMF magnitude at the leading edges of the high-speed solar wind streams are also seen,
indicating the CIR formed in the solar wind. Similar structures in the IMF and solar wind are observed over six solar rotations following CR2043.

Figures 5d-f display the GSE components of ${\bf G_{\perp}}(t)+{\bf G_{\parallel}}(t)$ calculated with $\alpha_\parallel=7.2$ and $\alpha_\perp=0.36$. These values were selected rather ad hoc and serve primarily as illustration. Qualitative features
would remain similar for a wide range of reasonable parameters. Each component shows a large
fluctuation due to the fluctuation of the IMF orientation ($b_x$, $b_y$, $b_z$) used in (12) and (19). To show the systematic variation of each component we filter this
fluctuation with a central 23-hour moving average of hourly data, plotted as the black curve. This systematic variation is dominated by the contributions from ${\bf G_{\perp}}(t)$,
the contribution from ${\bf G_{\parallel}}(t)$ being less than 20\% on average. The following correlations with IMF sector structure are apparent. First, $G_{z}(t)$ in Figure 5f tends to
be negative (positive) in T (A) sectors. This relationship between the sign of $G_{z}(t)$ and the IMF sector polarity is schematically shown in Figure 6. During $A<0$ epochs the
IMF points away from the Sun below the HCS (i.e. on the southern side of the HCS). The positive (negative) $G_{z}(t)$ in A (T) sectors indicates that the gradient is northward
below the HCS (Figure 6a), while it is southward above the HCS (Figure 6b). This implies that the GCR density distribution has a local maximum at the HCS. This is in a qualitative
agreement with drift model predictions for $A<0$ epochs. By analyzing the data observed with ground based detectors such as NMs and muon detectors, the bidirectional NS density
gradient has previously only been examined for drift effects on a yearly basis. The GMDN successfully observed it as a dynamic variation of the density gradient within a solar
rotation.

We next note, in Figures 5d \& e, that the ecliptic component $G_{x}(t)$ remains negative most of the time and shows enhancements following HCS crossings on DOYs 137 and 150. Such
enhancements are also seen in $G_{y}(t)$. The negative $G_{x}(t)$ implies that the ambient radial density gradient is positive and the density increases away from the Sun in
response to the outward convection. As shown in Figure 7, the enhancement of the ecliptic component of ${\bf G}(t)$ can be interpreted in terms of the bidirectional density
gradient toward the HCS ($G_{HCS}$) in the ecliptic plane, together with the contribution from the ambient radial gradient away from the Sun ($G_{bg}$). In this situation,
$G_{HCS}$ observed at the Earth is opposite to $G_{bg}$ before the HCS crossing (Figure 7a), while $G_{HCS}$ and $G_{bg}$ point in the same direction after the crossing (Figure 7b).
The resultant density gradient ($G_{net}$) observed at the Earth thus increases after the HCS crossing. This is observed as enhancements of $G_{x}(t)$ and $G_{y}(t)$ in
association with HCS crossings. These enhancements, therefore, also imply a local maximum of the GCR density at the HCS. We finally demonstrate that the systematic variations
discussed above are mainly due to drift streaming. Gray curves in Figure 8 show the three components of ${\bf G}(t)$ calculated by (20) for $\alpha_\perp = 0$. Note that
contributions from the parallel and cross-field diffusions are disregarded in this case. Nevertheless, the variations of $G_{x}(t)$ and $G_{y}(t)$ are not significantly different
from those for $\alpha_\perp = 0.36$ and $\alpha_\parallel= 7.2$, shown by black curves. It follows that the observed systematic variations in ${\bf G}(t)$ are predominantly caused
by drift streaming.

\subsection{Comparison with a simple drift model}\label{bozomath}

In this subsection, we compare the observed gradients with the numerical prediction of a simple drift model. The GCR density distribution is calculated by numerically solving Parker's
diffusive transport equation \citep{Par65} in a three-dimensional model heliosphere. The model incorporates all the important physical processes: anisotropic
diffusion, convection, particle drifts and adiabatic cooling in the expanding solar wind, as well as acceleration at shocks or in compression regions. The code assumes
steady-state conditions in the frame corotating with the Sun. The code is able to handle complex shapes of the current sheet and non-uniform solar wind speed leading to the
formation of CIRs. In this work, for brevity, we take a simple constant 400 km/s solar wind, hence the effects of CIRs are not considered. We adopt a slightly warped current
sheet, which leads to an asymmetric two-sector field. The tilt of the sheet is $\sim$15$\degr$, consistent with the average tilt angle of 12.6$\degr$ computed from the
photospheric magnetic field by the radial field model for CR2043 (see  http://wso.stanford.edu). The areas above and below the current sheet are equal, the radial magnetic field
component is set uniformly at 3.5 nT at 1 AU, while the total field in the helioequator at 1 AU (near the Earth) is 5 nT.

The code has been updated to calculate GCR gradients and anisotropies (see \cite{Kot01}). At high energies (tens of GeVs) the relative energy change of GCR in the heliosphere is
small, hence the spectral exponent remains equal to its interstellar value, and can be taken as a prescribed constant. The important parameters to determine the anisotropies are
the ratio of the three  components of the diffusion tensor,
\begin{equation}
\kappa_{\parallel} \, : \, \kappa_{A} \, : \, \kappa_{\perp} =
\alpha_{\parallel} \, : \, 1 \, : \, \alpha_{\perp}.
\end{equation}
Assuming that all these elements of the diffusion tensor scale inversely with the magnetic field strength, B, and are proportional to rigidity, these ratios remain constant.  In
this case, gradients will scale inversely  proportional to the particle rigidity while anisotropies remain constant, as long as the diffusive description is appropriate. The
diffusive approach becomes inaccurate and should break down at some limiting rigidity; the transition between diffusive and non-diffusive behaviour is not yet well understood
\citep{Kot99}.

In the model simulation used in this work, we adopt $\kappa_{\parallel}=2.4R_{L}c$ and $\kappa_{\perp}=0.12R_{L}c$, corresponding to $\alpha_{\parallel}=7.2$ and
$\alpha_{\perp}=0.36$. Comparison with more complex models including four-sector fields and CIRs are in progress and will be addressed in future work.

Figure 9 shows three components of the calculated density gradient of 50 GeV GCRs as measured at the Earth over one solar rotation period during a $A<0$ epoch. In each panel of
Figures 9b-d, open and filled circles connected by thin lines display the GSE component of the total gradient ${\bf G_{\perp}}(t)+{\bf G_{\parallel}}(t)$, while contributions from
${\bf G_{\perp}}(t)$ and ${\bf G_{\parallel}}(t)$ are shown by black and gray curves respectively. Although the calculated contribution from ${\bf G_{\parallel}}(t)$ is somewhat
larger, it is apparent that the temporal variation of the total gradient is mainly due to ${\bf G_{\perp}}(t)$, as in Figure 5. The overall magnitude of the calculated gradient is
also in reasonable agreement with the observations, except in the periods of observed strong enhancements following HCS crossings. The observed systematic variations in the
three components of the gradient are well reproduced by the simple drift model. First, $G_z(t)$ is negative (positive) in T (A) sectors (see Figure 9d). Second, the ecliptic
component $G_{x}(t)$ remains negative during the entire rotation period and shows enhancements after HCS crossings. Similar enhancements are also seen in $G_{y}(t)$ (see Figures
9b-c). These similarities between the observations and the model support drifts being responsible for the observed modulation of GCRs in the sample rotation period. In the next
section we discuss the similarity between the observed and predicted gradients in more detail.

In Figure 5 $G_{z}(t)$ is negative (positive) in T (A) sectors. On the other hand, $G_{x}(t)$ and $G_{y}(t)$ show these features in only two sectors starting from DOYs 137 and
151, but not in other two sectors starting from DOYs 130 and 145. There is no enhancement in the solar wind speed and IMF after DOY 145, while there are clear enhancements after
DOY 130. We currently have no clear interpretion of this. It also appears that $G_{x}(t)$ and $G_{y}(t)$ only show the above mentioned feature after HCS crossings from T-sectors
to A-sectors. If this is the case, it may be due to some asymmetry between the northern (T) and southern (A) hemispheres that is not taken into account in the present simple
drift model.

\section{SUMMARY AND DISCUSSION}

We derived the GCR density gradient in three dimensions from the anisotropy observed with the GMDN that has operated since March 2006. In particular, we have developed a new
analysis method for deducing the anisotropy free from atmospheric temperature effects on the muon intensity recorded by ground based detectors. The new method, together with the
global sky coverage available in the GMDN, allows us to accurately derive the anisotropy in three dimensions, including the north-south anisotropy, which has previously been
difficult. We analyze the anisotropy observed in 13 solar rotations between 2006 March 14 and 2007 March 4. During this period around solar activity minimum, clear signatures of
recurrent CIRs were seen with enhancement of the IMF magnitude at the leading edge of high-speed solar wind streams from coronal holes on the solar surface.

Inversely solving the diffusion flux equation for the spatial gradient of the GCR density, we deduced the gradient in three dimensions from the observed anisotropy corrected for both
the solar wind convection and the Compton-Getting effect arising from the Earth's orbital motion. By analyzing the gradient derived in a typical sample rotation (CR2043), we find
that the GSE-z component $G_z(t)$ of the gradient is negative (positive) in T (A) sectors. On the other hand, the GSE-x component $G_x(t)$ remains negative most of the time,
indicating the ambient radial gradient of the GCR density increased away from the Sun. Significant enhancements are also seen in both $G_x(t)$ and $G_y(t)$ following HCS
crossings. These features are seen in both the sample rotation and in the average of the 13 rotations observed. Statistical analyses of multiple rotations will be presented in a
separate paper. We demonstrate that these features can be interpreted in terms of the local maximum of GCR density at the HCS which is consistent with the prediction of the drift
model for $A<0$ epochs. It is noted from Figure 8 that similar features are also seen in the gradient derived when neglecting the contribution from the parallel and perpendicular
diffusion. This confirms that the observed anisotropy is dominated by drift streaming at these high energies.

We also compare these features with the numerical solution of the GCR transport equation in the three dimensional heliosphere with a simple two sector IMF structure. There is an
overall resemblance between the observed gradient and the prediction by this simple model, ignoring the CIR signatures such as strong enhancements of both the solar wind velocity
and the IMF magnitude. This comparison with the model is preliminary. The model does not take account of effects of CIRs. In particular, the enhanced IMF magnitude in the CIR leads to a
reduced Larmor radius $R_{L}(t)$ and an enhanced gradient by a factor of $1/R_{L}(t)$ as seen in (19) and (20). By comparing gradients derived employing an IMF with and without
this magnitude enhancement, we confirmed this effect contributed about 50 \% of the enhanced gradient. The remaining half of the enhancement still exists, even without the IMF
enhancement, with temporal variations similar to those above. The model we used in this work also assumes a two-sector IMF structure, while the observed sample rotation clearly
shows the four-sector structure. The average features of the gradient mentioned above are observed in only two sectors, but are not observed in the other sectors. Further analyses
based on a more realistic model are needed to clarify this.

The discussions above suggest that the GCR modulation by CIRs does not organize the observed gradient, but the drift effect in the large scale magnetic field plays a more
important role. This seems to contradict the conclusion, derived from satellite measurements of sub-GeV particles, that the GCR modulation is not organized by the HCS
\citep{Ric04}. A possible explanation for this discrepancy is the difference in particle energy. Sub-GeV GCRs have smaller Larmor radii and are more sensitive to small-scale
structures in the solar wind. Conversely, the primary GCRs to which GMDN network responds are much more energetic ($\sim$ 50-150 GeV) and less sensitive to small-scale structures,
selectively responding to the larger scales over which the global drift mechanism operates. It is also reported in a recent paper 
by \cite{Min07} that drifts are suppressed by magnetic turbulence, but the suppression sets in at lower turbulence amplitudes for low-energy cosmic rays than for high-energy cosmic rays. This may give a possible explanation for why drift effects might be dominant at muon detector energies, but negligible at sub-GeV energies. The difference in the analysis method may also be responsible for the discrepancy.
Analyses reported by \cite{Ric04} were based on a one dimensional spatial distribution of the GCR density along the viewing path of a detector moving through the solar wind. Our analysis,
on the other hand, uses the GCR spatial gradient in three dimensions, which provides us with additional information on the distribution around the detectors' viewing paths. It is
not always easy to deduce the large-scale distribution of GCRs solely from the temporal variation of the density, particularly when the amplitude of the variation is small. The
modulation of the GCR density in the CIR observed during the solar minimum period is generally small, when compared with other violent events such as the Forbush decrease caused
by the Earth's encounter with a CME accompanied by a strong shock. This is seen in the best-fit density in Figure 4a. The modulation of the density in this figure is less than 1
\% with no clear correlation with HCS crossings and/or the CIR structures in Figures 5a-c.

In conclusion, it is demonstrated that both the observation of high-energy GCRs with the GMDN and employing a new analysis method provides us with new information about the
density gradient in three dimensions. In particular we found that the gradient observed in a sample solar rotation period shows systematic variations correlated with the Earth's
HCS crossings. These features are in good agreement with the prediction of a simple drift model for $A<0$ epochs. However, they are only seen in two of the four sectors observed
in the sample rotation period. Clarifying the physical mechanism responsible for this requires further analyses, including statistical analysis of multiple rotations and detailed
comparisons using more realistic model calculations. Analysis of the gradient observed by the GMDN in association with CIR structures in $A>0$ epochs will give us a unique
opportunity to test the drift model.



\acknowledgments

This work is supported in part by U.S. NSF grant ATM-0527878 and NASA grant NNX 07-AH73G, and in part by Grants-in-Aid for Scientific Research from the Ministry of Education, Culture, Sports,
Science and Technology in Japan and by the joint research program of the Solar-Terrestrial Environment Laboratory, Nagoya University. The observations with the Kuwait University
muon detector are supported by the Kuwait University Research grant SP03/03. We thank N. F. Ness for providing {\it{ACE}} magnetic field data via the {\it{ACE}} Science Center.

\appendix

\section{Appendix: Atmospheric temperature effect on the muon intensity}

The fractional deviation of the muon intensity $I(x, \theta, E_{\mu}, t )$ from its average due to the atmospheric temperature effect at a time $ t $, zenith angle $ \theta $,
atmospheric depth $ x $ and muon threshold energy $ E_{\mu} $ is given as
\begin{equation}
\Delta I/I(x, \theta, E_{\mu}, t ) =  \int_{0}^{x} \alpha (x', \theta, E_{\mu}) \delta T(x', t) dx',
\end{equation}
where $ \delta T(x', t) $ is the temperature deviation from the average at an atmospheric depth $ x' $ and $ \alpha (x', \theta, E_{\mu} ) $ is the partial temperature coefficient
\citep{Sag86}. The muon intensity variation due to the temperature effect can be derived from this equation, when $ \delta T(x', t) $ is available at various $ x' $. Accurate 
$ \delta T(x', t) $ are seldom available for all $ x' $ and all muon detectors simultaneously. In this paper, we use an alternative correction for temperature effects. As described
briefly in section 1, the negative temperature effect on the muon intensity measured with the surface-level detectors arises predominantly from the increase of muon decays arising
from the increased path length caused by atmospheric expansion. A significant negative correlation, therefore, is expected between the altitude of the equi-pressure surface and
the muon intensity; i.e., the muon intensity decreases with increasing this altitude due to atmospheric expansion.

Figure 10a shows the percent deviation of the pressure corrected hourly count rate of muons recorded in the vertical channel at Nagoya from its yearly mean in 2006, set to 100 \%.
The observed muon rate at Nagoya (gray curve) shows a seasonal variation with a maximum in the northern hemisphere winter. We examine the temperature effect on the muon rate at
Nagoya by analyzing high-altitude atmospheric data available from Shionomisaki (at geographical latitude and longitude of 33.5$\degr$N, 135.8$\degr$E) located south-west of
Nagoya. These radio-sonde measurement have been continuously conducted by the Japan Meteorological Agency every 12 hours. The thin black curve in Figure 10a displays the 100 hPa
altitude equi-pressure surface (henceforth called ``the 100 hPa altitude"). The anti-correlation between the muon rate and the 100 hPa altitude is evident. The anti-correlation is
confirmed in Figure 10b showing the scatter plot between the two quantities, with a correlation coefficient of -0.947 and a regression coefficient of -6.83 \%/km. The muon rate
corrected using the 100 hPa altitude and this regression coefficient is shown by the thick black curve in Figure 10a. The seasonal variation in the gray curve is eliminated by
this correction. Similar correlations are seen in Figure 10c for the 100-300 hPa equi-pressure range. In this paper, we use the 100 hPa altitude and the regression coefficient to
correct the temperature effect on the hourly muon rate in each directional channel in the GMDN. For this correction, we calculate the hourly value of the 100 hPa altitude by
interpolating the data observed twice a day. We use the 100 hPa altitude, also measured twice a day, at Port Alegre (30.0$\degr$S, 308.8$\degr$E), Hobart (42.8$\degr$S,
147.5$\degr$E) and Kuwait (29.2$\degr$N, 48.0$\degr$E) airports for corrections of the muon rates recorded by S\~{a}o Martinho, Hobart and Kuwait University muon detectors,
respectively.\footnote{The high-altitude atmospheric data are available at http://badc.nerc.ac.uk/home/} The distance between each muon detector and its associated
meteorological station is $\sim$200 km for Shionimisaki, $\sim$260 km for Port Alegre, $\sim$30 km for Hobart and $\sim$10 km for Kuwait. The number of 100 hPa altitude
observations available, from twice daily measurements over the 13 rotation periods analyzed, is 693 from Shionomisaki, 576 from Hobart, 271 from Porto Alegre and 492 from Kuwait.
The correlation ($\gamma_{temp}$) and regression ($\beta_{temp}$) coefficients obtained for all directional channels in the GMDN are listed in the last two columns of Table 1.
Note that we cannot correct the GMDN data routinely with $\beta_{temp}$ in this table, because the 100 hPa altitude data are not always available for all muon detectors. We could
make such corrections for only 79 of the $\sim$350 days analyzed here.

The correlations for Nagoya and Kuwait University are very significant with $\gamma_{temp}$ less than -0.9, whilst those for Hobart and S\~{a}o Martinho are less significant, with
$\gamma_{temp}$ ranging from -0.773 to -0.735. This is mainly due to the amplitude of the seasonal variation of the 100 hPa altitude being different for the different locations.
The correlation coefficient is high for Nagoya and Kuwait University where the amplitudes are as large as 0.268 km and 0.265 km, respectively, whilst it is lower for Hobart and
S\~{a}o Martinho where the amplitudes are only 0.159 km and 0.071 km, respectively. It is noted in Table 1 that the $\beta_{temp}$ values are almost identical for all directional channels
for the Nagoya and Kuwait University detectors, where the correlation is very significant. The fractional difference between the maximum and minimum $\beta_{temp}$ is only 14 \%
for Nagoya and 20 \% for Kuwait University. It is also seen for Nagoya that $\beta_{temp}$ is minimum for the vertical channel and gradually increases for more inclined channels.
This confirms the expectation, on which we based the new analysis method for eliminating the temperature effect, that the effect is similar for all directional channels, at least
as a first order approximation (see section 3 in the text).

\clearpage

\begin{deluxetable}{ccccrrrr}
\tabletypesize{\footnotesize}
\tablewidth{0pt}
\tablecaption{Characteristics of the Global Muon Detector Network}
\tablehead{
\colhead{muon detector (geographical location) } & \colhead{count rate}        & \colhead{$\sigma_c$}     &
\colhead{$P_m$}                                 & \colhead{$\lambda_{asymp}$} & \colhead{$\phi_{asymp}$} &
\colhead{$\gamma_{temp}$}                       & \colhead{$\beta_{temp}$}                                   \\
\colhead{directional channel }                  & \colhead{[$10^4/h$]}        & \colhead{[\%]}           &
\colhead{[GV]}                                  & \colhead{$[\degr]$}         & \colhead{$[\degr]$}      &
\colhead{                                                                   } & \colhead{[\%/km]}            \\
} \startdata
Nagoya (35.1$\degr$N, 137.0$\degr$E, 77 m) \\
V  & 276 & 0.06 & 59.4 & 28.0 & 168.4 & -0.947 & -6.83 \\
N  & 125 & 0.09 & 64.6 & 47.0 & 192.7 & -0.950 & -7.11 \\
S  & 123 & 0.10 & 62.6 &  2.9 & 157.5 & -0.948 & -7.10 \\
E  & 120 & 0.09 & 66.5 & 10.8 & 194.0 & -0.950 & -7.08 \\
W  & 126 & 0.09 & 61.8 & 40.2 & 135.0 & -0.949 & -7.16 \\
NE &  58 & 0.14 & 72.0 & 25.9 & 209.4 & -0.953 & -7.29 \\
NW &  62 & 0.13 & 66.6 & 64.3 & 156.4 & -0.952 & -7.37 \\
SE &  58 & 0.14 & 69.3 & -6.6 & 182.4 & -0.953 & -7.37 \\
SW &  60 & 0.13 & 65.6 & 12.8 & 131.1 & -0.949 & -7.35 \\
N2 &  61 & 0.13 & 83.0 & 56.1 & 217.0 & -0.960 & -7.66 \\
S2 &  60 & 0.13 & 80.5 &-14.1 & 152.2 & -0.958 & -7.60 \\
E2 &  58 & 0.14 & 88.3 &  2.0 & 206.8 & -0.959 & -7.51 \\
W2 &  62 & 0.13 & 79.3 & 40.4 & 105.0 & -0.958 & -7.54 \\
N3 &  18 & 0.27 &105.0 & 59.5 & 236.1 & -0.945 & -7.77 \\
S3 &  18 & 0.27 &103.7 &-24.4 & 149.6 & -0.941 & -7.79 \\
E3 &  17 & 0.28 &113.7 & -1.7 & 213.4 & -0.938 & -7.61 \\
W3 &  18 & 0.27 &103.0 & 35.6 &  87.5 & -0.945 & -7.59 \\
\tableline
Hobart (43.0$\degr$S, 147.3$\degr$E, 65 m) \\
V  &  83 & 0.12 & 54.6 &-40.0 & 170.5 & -0.773 & -4.44 \\
N  &  29 & 0.20 & 59.0 &-17.5 & 152.5 & -0.767 & -4.47 \\
S  &  30 & 0.20 & 59.0 &-53.9 & 206.0 & -0.783 & -4.62 \\
E  &  30 & 0.20 & 59.0 &-18.6 & 193.5 & -0.749 & -4.57 \\
W  &  29 & 0.20 & 59.0 &-55.6 & 132.4 & -0.770 & -4.51 \\
NE &  12 & 0.33 & 63.7 & -3.9 & 176.0 & -0.696 & -4.46 \\
NW &  11 & 0.33 & 63.7 &-29.1 & 125.8 & -0.716 & -4.31 \\
SE &  12 & 0.33 & 63.7 &-30.2 & 214.8 & -0.696 & -4.31 \\
SW &  12 & 0.33 & 63.7 &-77.2 & 171.6 & -0.756 & -4.54 \\
N2 & 7.0 & 0.42 & 76.7 &  0.2 & 145.6 & -0.695 & -4.88 \\
S2 & 7.3 & 0.42 & 76.3 &-57.6 & 236.7 & -0.698 & -4.64 \\
E2 & 7.2 & 0.42 & 76.3 & -6.4 & 205.8 & -0.649 & -4.39 \\
W2 & 7.1 & 0.42 & 76.3 &-53.2 &  95.6 & -0.735 & -7.56 \\
\tableline
S\~{a}o Martinho (29.4$\degr$S, 306.2$\degr$E, 488 m) \\
V  & 231 & 0.07 & 55.6 &-22.6 & 330.4 & -0.735 & -4.18 \\
N  &  88 & 0.11 & 59.8 &  5.3 & 325.3 & -0.680 & -3.83 \\
S  &  91 & 0.11 & 59.1 &-48.3 & 347.2 & -0.734 & -4.37 \\
E  & 102 & 0.10 & 61.7 &-10.7 & 358.5 & -0.711 & -3.96 \\
W  & 102 & 0.10 & 58.3 &-29.1 & 298.0 & -0.713 & -4.02 \\
NE &  42 & 0.15 & 66.6 & 10.3 & 350.3 & -0.665 & -3.67 \\
NW &  42 & 0.15 & 62.7 & -0.9 & 299.0 & -0.653 & -3.71 \\
SE &  43 & 0.15 & 65.2 &-30.6 &  11.2 & -0.688 & -4.05 \\
SW &  43 & 0.15 & 62.3 &-56.8 & 304.0 & -0.710 & -4.01 \\
N2 &  29 & 0.17 & 79.0 & 23.0 & 322.5 & -0.636 & -3.45 \\
S2 &  30 & 0.17 & 77.3 &-63.1 &   8.8 & -0.692 & -4.09 \\
E2 &  37 & 0.15 & 80.6 & -3.6 &  12.9 & -0.638 & -3.66 \\
W2 &  37 & 0.15 & 75.0 &-27.7 & 273.0 & -0.707 & -4.01 \\
N3 & 3.4 & 0.46 & 99.0 & 33.3 & 321.5 & -0.494 & -3.69 \\
S3 & 3.5 & 0.46 & 96.9 &-68.6 &  32.6 & -0.589 & -5.11 \\
E3 & 7.5 & 0.30 &105.0 & -0.7 &  20.0 & -0.637 & -4.58 \\
W3 & 7.7 & 0.30 & 98.8 &-23.7 & 257.9 & -0.654 & -4.39 \\
\tableline
Kuwait-University (29.3$\degr$N, 48.0$\degr$E, 0 m) \\
V  &  86 & 0.10 & 62.3 & 24.2 &  77.2 & -0.945 & -6.43 \\
N  &  22 & 0.19 & 67.8 & 61.4 &  76.3 & -0.948 & -6.55 \\
S  &  22 & 0.19 & 69.4 &-11.9 &  90.1 & -0.948 & -6.85 \\
E  &  22 & 0.20 & 73.5 & 18.1 & 121.1 & -0.950 & -6.36 \\
W  &  22 & 0.19 & 66.0 & 12.6 &  36.4 & -0.945 & -6.73 \\
NE & 6.4 & 0.35 & 78.2 & 44.4 & 127.9 & -0.946 & -6.55 \\
NW & 6.5 & 0.34 & 72.9 & 45.1 &  18.8 & -0.942 & -6.50 \\
SE & 6.5 & 0.35 & 82.3 & -6.0 & 119.7 & -0.946 & -6.35 \\
SW & 6.8 & 0.34 & 73.4 &-17.4 &  56.1 & -0.939 & -7.10 \\
N2 & 2.8 & 0.47 & 97.9 & 81.9 &  58.9 & -0.916 & -6.14 \\
S2 & 2.8 & 0.47 &102.2 &-26.1 &  95.9 & -0.908 & -5.94 \\
E2 & 2.6 & 0.49 &109.8 & 15.9 & 135.9 & -0.908 & -6.38 \\
W2 & 2.8 & 0.47 & 97.0 &  0.8 &  15.8 & -0.927 & -6.75 \\
\enddata

\tablenotetext{a}{Average hourly count rate, count rate error ($\sigma_c$), median primary rigidity ($P_m$), latitude ($\lambda_{asymp}$) and longitude ($\phi_{asymp}$),
correlation ($\gamma_{temp}$) and regression ($\beta_{temp}$) coefficients of the correlation between the 100 hPa altitude and the muon rate are listed for each GMDN directional
channel. See Appendix A for $\gamma_{temp}$ and $\beta_{temp}$.} \tablenotetext{b}{The Kuwait University muon detector uses proportional counters; the other detectors use plastic
scintillation counters (see text).}


\end{deluxetable}

\clearpage

\begin{figure}
\plotone{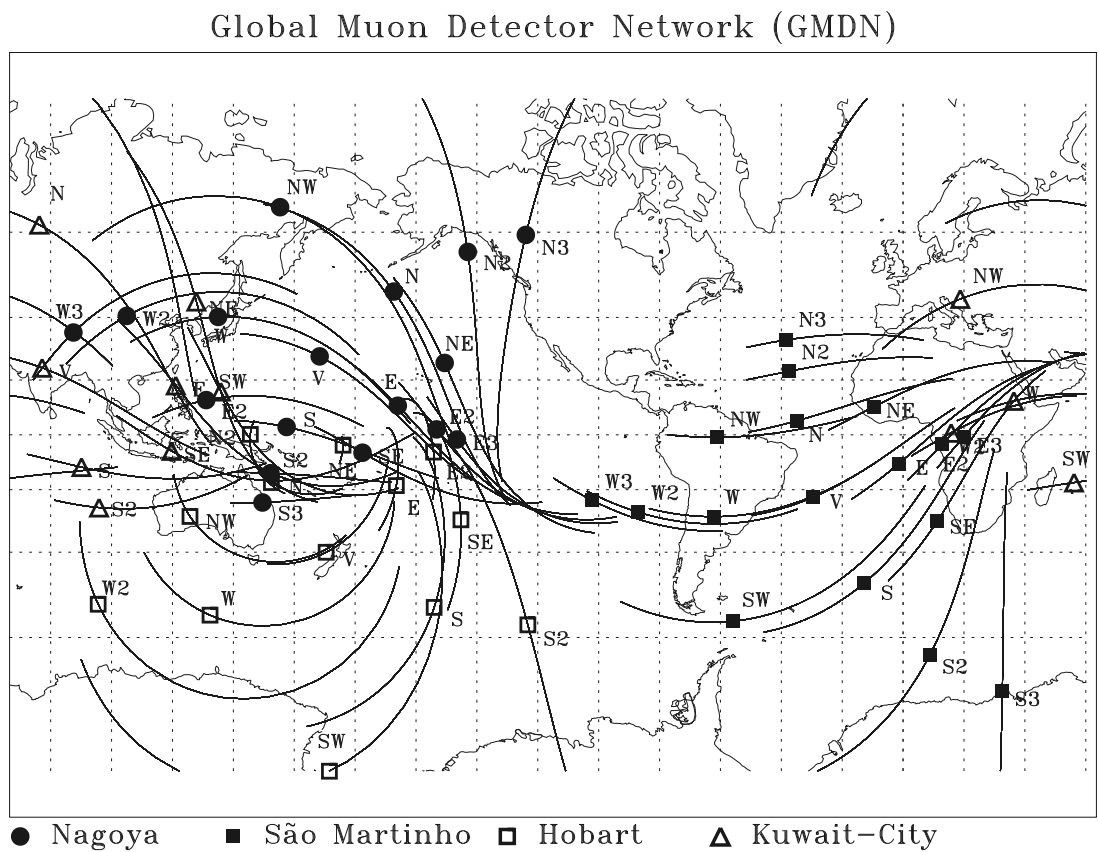}
\caption{Asymptotic viewing directions of the GMDN. Each symbol (filled circles for Nagoya, filled squares for S\~{a}o Martinho, open squares for Hobart and open
triangles for Kuwait University) shows the asymptotic viewing direction (after correction for geomagnetic bending) of each directional channel with median primary rigidity ($P_m$)
as listed in Table 1. The track through each symbol represents the spread of viewing directions corresponding to the central 80 \% of each channel's energy response (see text).}
\end{figure}

\clearpage

\begin{figure}
\epsscale{.80}
\plotone{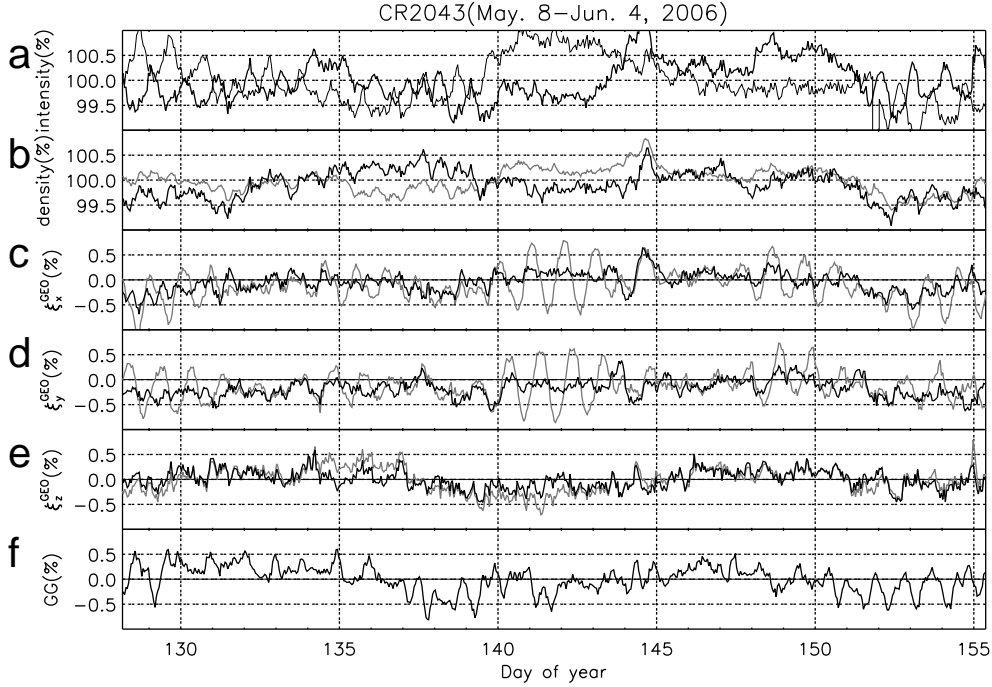}
\caption{Best-fit parameters in equation (1) derived from best-fits to the hourly GMDN data in a sample solar rotation period, CR2043 (May 8 to June 4, 2006). From top to bottom,
(a) the pressure corrected vertical muon rates, (b) the best-fit density, (c-e) the three components of anisotropy in the GEO and (f) the Nagoya GG component are shown plotted
against time. The black and thin black curves in (a) show respectively the vertical muon rates at Nagoya and S\~{a}o Martinho. The gray curves in (b-e) display parameters obtained
from best-fitting to the original muon rate, while the black curves in (c-e) show parameters by the new analysis method. Note that the best-fit density in (b) cannot be directly
derived from the new analysis method. The density by the new method displayed in (b) is derived by using the vertical muon rate at Nagoya, corrected for the atmospheric
temperature effects (see text). The bottom panel (f) shows the 5-hour central moving average of the hourly value of the GG component.}
\end{figure}

\clearpage

\begin{figure}
\epsscale{.80}
\plotone{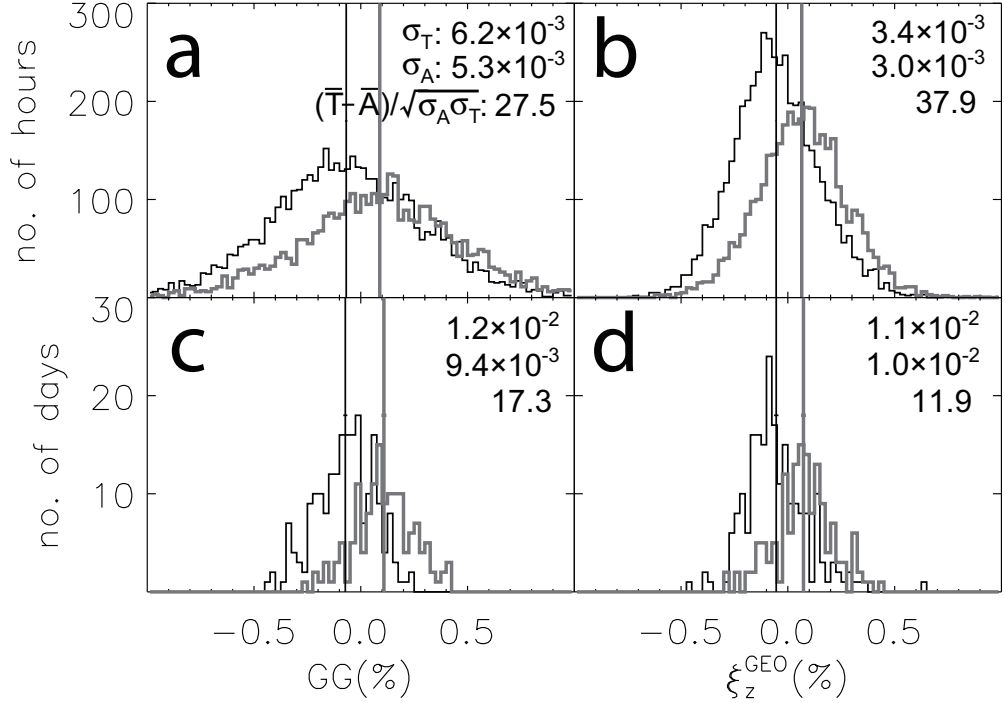}
\caption{Correlation of the NS anisotropy with the IMF sector polarity over 13 solar rotations CR2041-CR2053. The left two panels (a \& c)
display histograms of the GG component, whilst the right two panels (b \& d) show histograms of $ \xi^{GEO}_{z} $ by the new analysis method. The gray and black traces in each
panel show histograms in T and A sectors respectively. The upper two panels are for hourly values and the lower two panels are for daily mean values. Average values in T and A
sectors ($\bar{T}$ and $\bar{A}$) are indicated by vertical lines. Standard deviations of T and A sectors ($\sigma_T$ and $\sigma_A$ in \%) and statistical significances of T/A
separation defined as $(\bar{T}-\bar{A})/\sqrt{\sigma_{T}\sigma_{A}}$ are also indicated in each panel.}
\end{figure}

\clearpage

\begin{figure}
\epsscale{.80}
\plotone{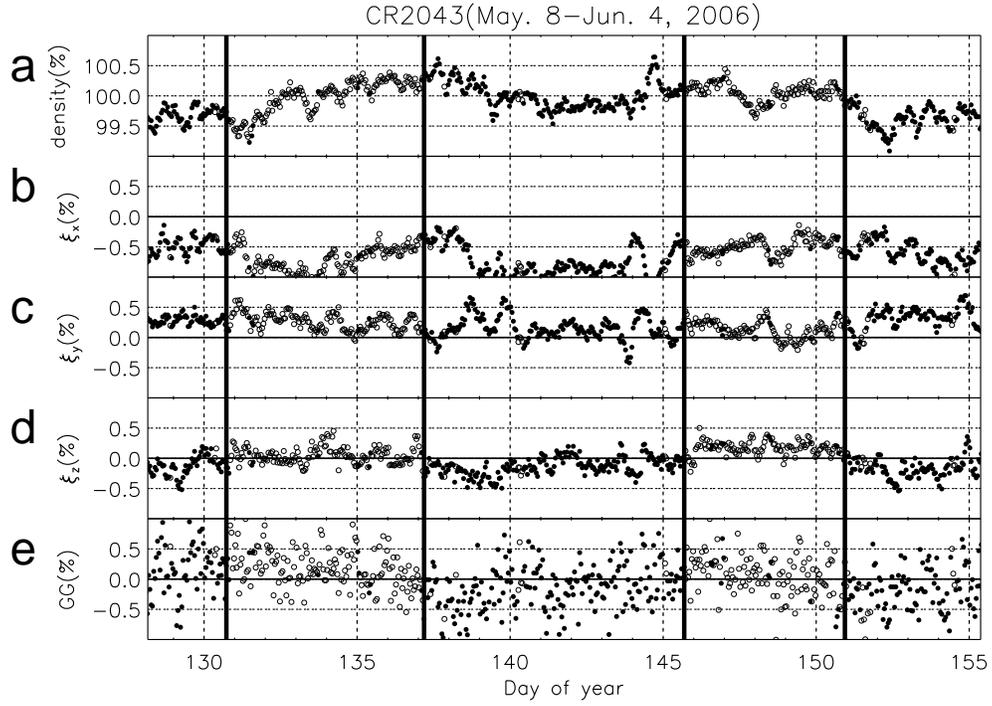}
\caption{Best-fit anisotropy components in GSE coordinates for CR2043. The anisotropy from Figure 2 (in GEO coordinates) is transformed to GSE
coordinates and corrected for solar wind convection and the Compton-Getting effect (see text). From top to bottom, each panel displays (a) the best-fit density, (b-d) the three
components of the anisotropy and (e) the hourly values of the GG component as a function of time (DOY). Open (filled) circles in each panel show the parameter in T (A) sector.
The time of the Earth's HCS crossing (as determined by the reversal of the IMF longitude) is indicated by the vertical line.}
\end{figure}

\clearpage

\begin{figure}
\epsscale{.80}
\plotone{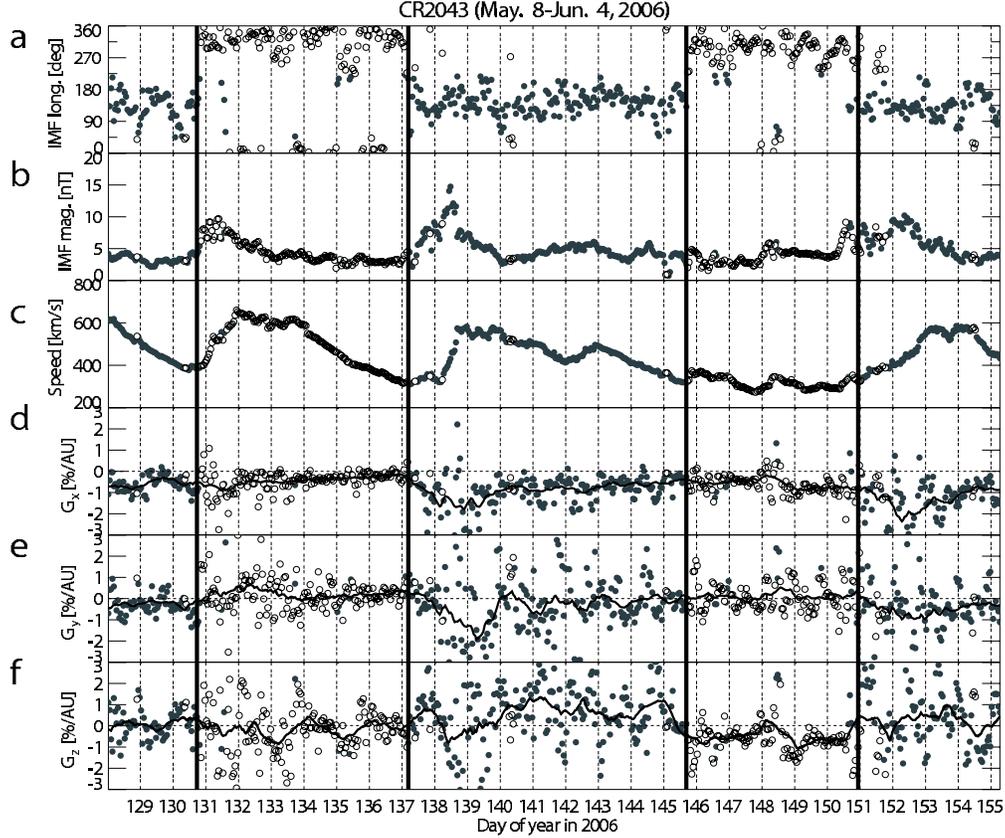}
\caption{Solar wind parameters and the GCR density gradient during CR2043: (a) Hourly mean GSE longitude of the IMF, (b) magnitude of the IMF,
(c) solar wind speed and three GSE components of the gradient, (d) $G_{x}$, (e) $G_{y}$ and (f) $G_{z}$ derived with $\alpha_\perp = 0.36$ and $\alpha_\parallel= 7.2$ (see text).
Open and filled circles display values in T and A sectors, respectively. Panels (d)-(f) display hourly values of the component of ${\bf G_{\perp}}(t)+{\bf G_{\parallel}}(t)$. Each
thin black curve in (d)-(f) displays the 23-hour central moving average of the hourly gradient and shows the systematic variation of the gradient by filtering the large scattering
of data points due to the fluctuation of the IMF orientation. The timing of each HCS crossing is indicated by the vertical line.}
\end{figure}

\clearpage

\begin{figure}
\epsscale{.60}
\plotone{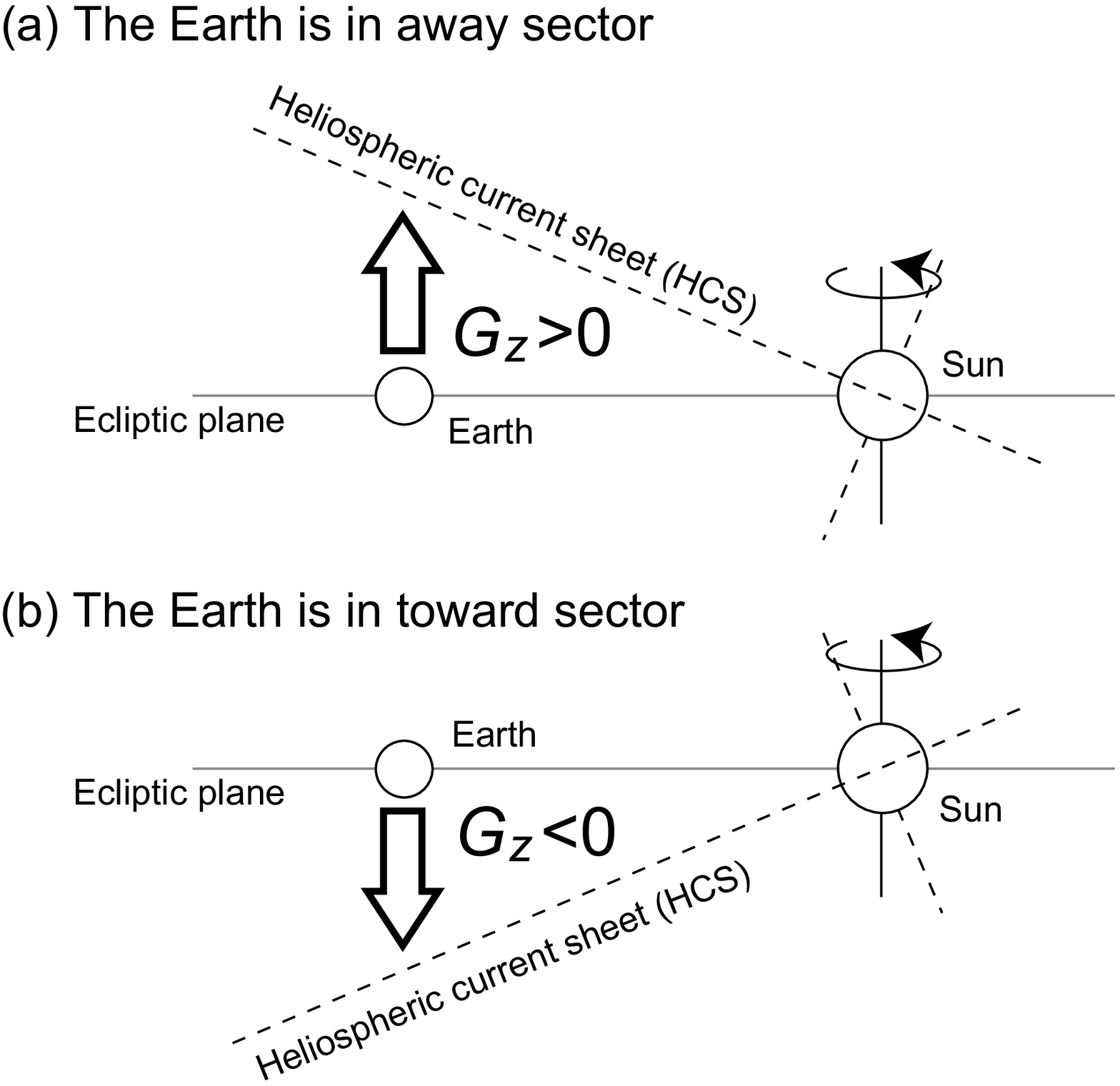}
\caption{Schematic view in the meridian plane of $G_{z}$ changing sign according to the IMF sector polarity. Panels (a) and (b) represent
situations when the Earth is on the southern side of the HCS (A sector) and when it is on the northern side (T sector) respectively.}
\end{figure}

\clearpage

\begin{figure}
\epsscale{.8}
\plotone{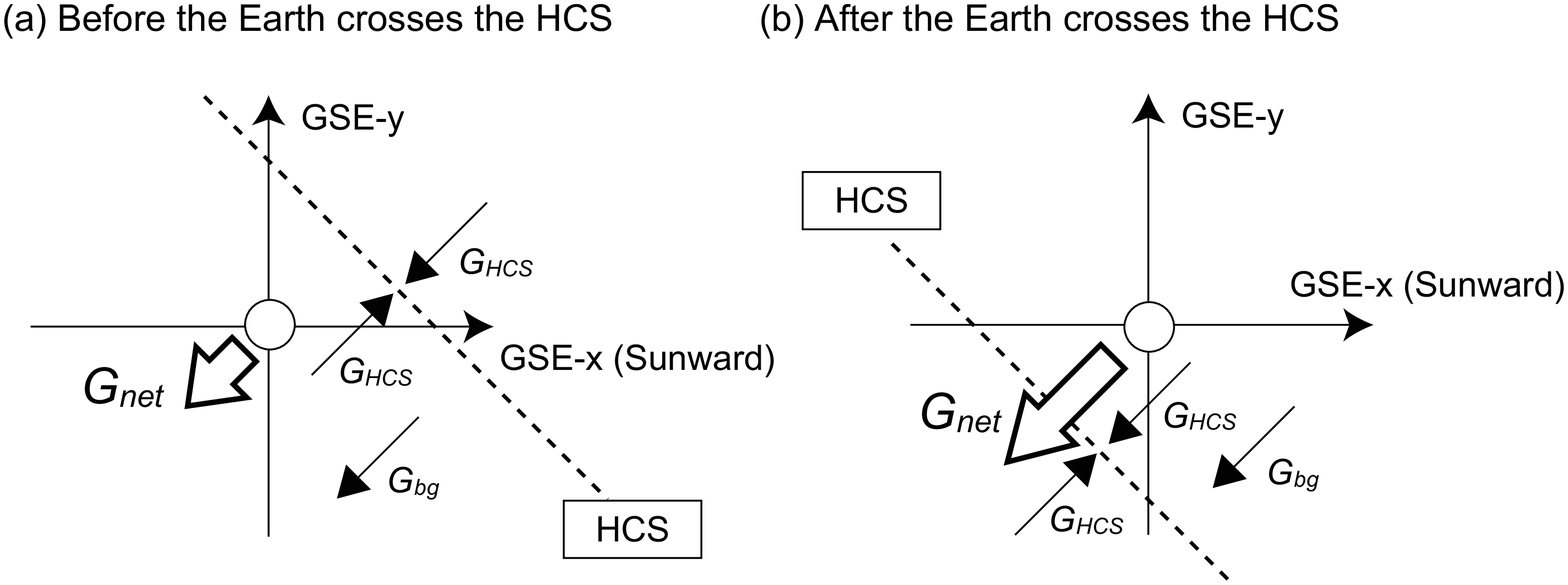}
\caption{Schematic view of the ecliptic component of $\bf{G}$ varying in association with the Earth's HCS crossing. Panels (a) and (b) represent
situations on the ecliptic plane before and after the HCS crossing, respectively. The figure displays only the gradient components that are perpendicular to the line of intersection of the HCS with the ecliptic plane.}
\end{figure}

\clearpage

\begin{figure}
\plotone{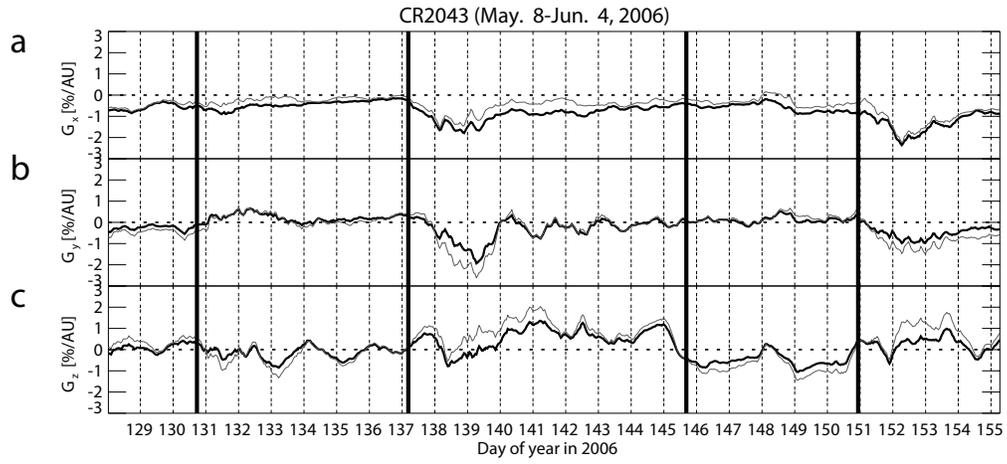}
\caption{Variations of the three GSE components of ${\bf G}$ for different $\alpha_\perp$ and $\alpha_\parallel$: (a) $G_{x}$, (b) $G_{y}$ and (c) $G_{z}$. In
each panel, the black line displays ${\bf G}(t)$ for $\alpha_\perp = 0.36 $ and $\alpha_\parallel= 7.2 $, while the gray line shows ${\bf G}(t)$ for $\alpha_\perp = 0$ and
{\boldmath{$\xi_\parallel$}}= 0 for the pure drift streaming case. The timing of each HCS crossing is indicated by the vertical line.}
\end{figure}

\clearpage

\begin{figure}
\plotone{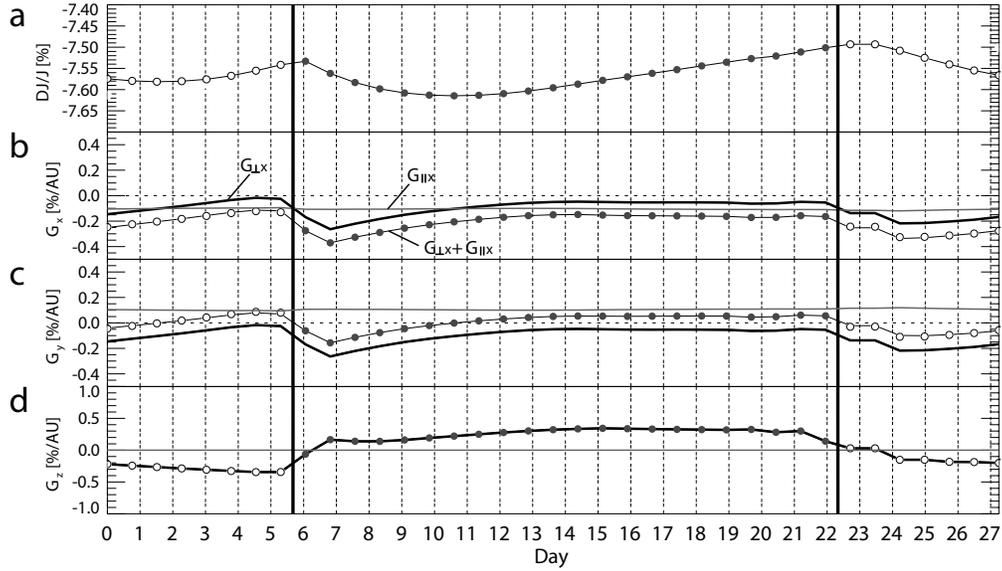}
\caption{27-day variation in the gradient, ${\bf G}$, of 50 GeV GCRs calculated numerically from a simple three dimensional drift model for $A < 0 $. The model
assumes $\alpha_\perp = 0.36 $ and $\alpha_\parallel= 7.2 $ and two IMF sectors separated by a wavy HCS with a tilt angle of $15^\circ$. No CIRs are included. From top to bottom,
each panel displays (a) the GCR density normalized to the interstellar value and (b-d) the three GSE components of $\bf{G}$. Open (filled) circles connected by a thin line in each
panel display the hourly values in T (A) sectors. In each panel (b)-(d), the black trace displays the component of ${\bf G_{\perp}}(t)$, while the gray trace shows the contribution of ${\bf G_{\parallel}}(t)$.}
\end{figure}

\clearpage

\begin{figure}
\epsscale{0.8}
\plotone{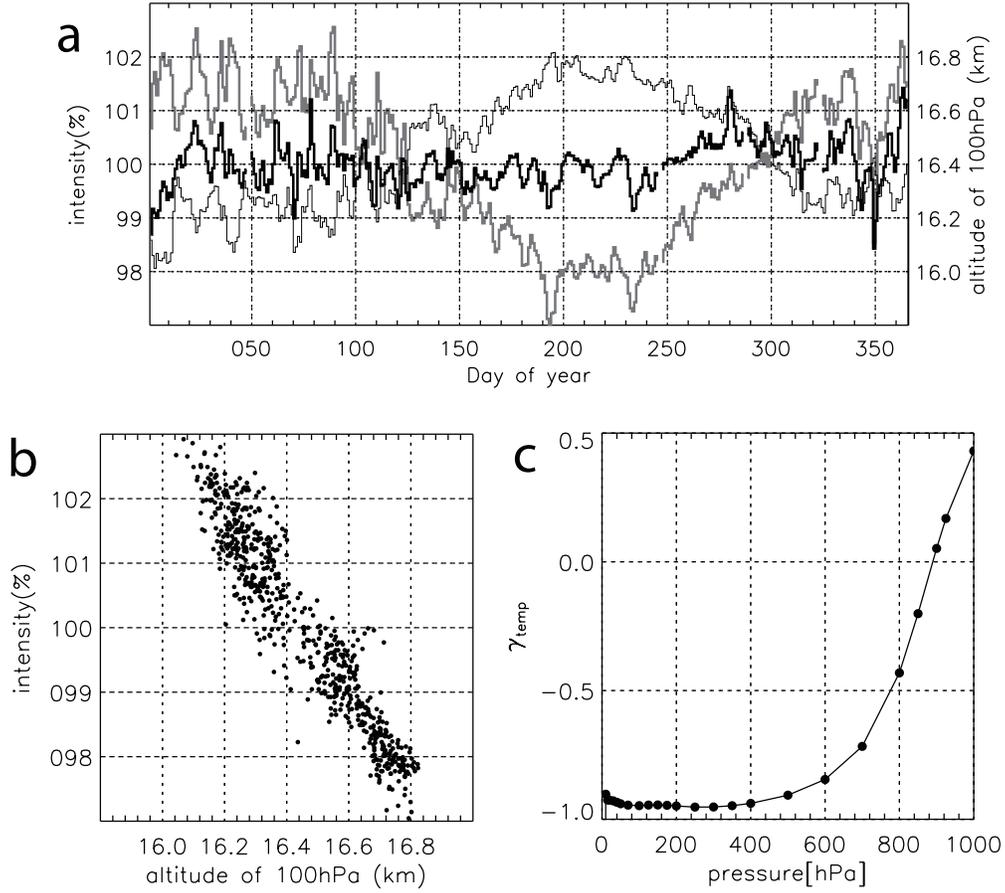}
\caption{Atmospheric temperature effects on the pressure-corrected muon count rate. The top panel (a) displays the observed vertical muon rates at Nagoya by a
gray trace (with the left axis), the 100 hPa altitude by the thin black trace (with the right axis) and the muon rate corrected for atmospheric temperature effects by the thick
black trace (with the left axis), each as a function of time (DOY) in 2006. The bottom left panel (b) displays a scatter plot of the observed muon rate and the 100 hPa altitude,
whilst the bottom right panel (c) shows the correlation coefficient between the observed muon rate and the altitude as a function of the atmospheric pressure of the equi-pressure
surface.}
\end{figure}

\end{document}